\documentclass[pdflatex,sn-mathphys]{sn-jnl}% Math and Physical Sciences Reference Style
\usepackage{physics}
\usepackage{amsmath}
\usepackage{graphicx}
\jyear{2021}%

\theoremstyle{thmstyleone}%
\theoremstyle{thmstyletwo}%

\theoremstyle{thmstylethree}%

\begin{document}

\title[A Theoretical Thrust Density Limit for Hall Thrusters]{A Theoretical Thrust Density Limit for Hall Thrusters}

%%=============================================================%%
%% Prefix	-> \pfx{Dr}
%% GivenName	-> \fnm{Joergen W.}
%% Particle	-> \spfx{van der} -> surname prefix
%% FamilyName	-> \sur{Ploeg}
%% Suffix	-> \sfx{IV}
%% NatureName	-> \tanm{Poet Laureate} -> Title after name
%% Degrees	-> \dgr{MSc, PhD}
%% \author*[1,2]{\pfx{Dr} \fnm{Joergen W.} \spfx{van der} \sur{Ploeg} \sfx{IV} \tanm{Poet Laureate} 
%%                 \dgr{MSc, PhD}}\email{iauthor@gmail.com}
%%=============================================================%%

\author*[1,2]{\fnm{Jacob} \sur{Simmonds}}\email{jacobbs@princeton.edu}

\author[2]{\fnm{Yevgeny} \sur{Raitses}}\email{yraitses@pppl.gov}
\equalcont{These authors contributed equally to this work.}

\author[3]{\fnm{Andrei} \sur{Smolyakov}}\email{andrei.smolyakov@usask.ca}
\equalcont{These authors contributed equally to this work.}

\affil*[1]{\orgdiv{Mechanical and Aerospace Engineering}, \orgname{Princeton University}, \orgaddress{\street{Princeton University}, \city{Princeton}, \postcode{08543}, \state{NJ}, \country{USA}}}

\affil[2]{\orgdiv{Plasma Sciences and Technology}, \orgname{Princeton Plasma Physics Laboratory}, \orgaddress{\street{100 Stellarator Road}, \city{Princeton}, \postcode{08544}, \state{NJ}, \country{USA}}}

\affil[3]{\orgdiv{Theoretical Plasma Physics}, \orgname{University of Saskatchewan}, \orgaddress{\street{116 Science Place}, \city{Saskatoon}, \postcode{S7N 5E2}, \state{SK}, \country{Canada}}}

%%==================================%%
%% sample for unstructured abstract %%
%%==================================%%

\abstract{Hall Thrusters typically operate at thrust densities on the order of 10 N/m$^2$, which is well below the thrust density limits discussed in previous literature. These limits have been considered here and each component of thrust density is analyzed to demonstrate the relative contribution to the total thrust density. Dependencies of the thrust density limits upon the thruster geometry, electron mobility, and the applied magnetic field are revealed and compared with experimental data.}


\maketitle

\section{Introduction}\label{sec1}

Hall thrusters are quickly maturing as a technology to become the most common form of electric propulsion in space.\cite{Proportion_Thrusters} Recent scientific missions such as PSYCHE are now utilizing a Hall thruster as the main propulsion unit for deep-space orbital transits and maneuvering,\cite{PsycheMission} and large constellations of commercial satellites such as Starlink rely upon Hall thrusters to maintain their orbits. These new developments are brought about by two converging factors in the space industry: the rise of higher power capabilities onboard satellites and the miniaturization of components. These two enabling developments call for thruster technologies that can provide high throughput at small physical scales. A quantitative measure of such a quantity is the thrust density of the thruster, which is simply the thrust generated divided by the front-facing area of the thruster. The interest and development of Hall thrusters is in part due to their favorable thrust density compared to other forms of electric propulsion such as gridded ion thruster in the 100 W - 10 kW power range, as well as their favorable thrust-specific impulse ratio.

\begin{figure}
    \centering
    \includegraphics[width=0.75\textwidth]{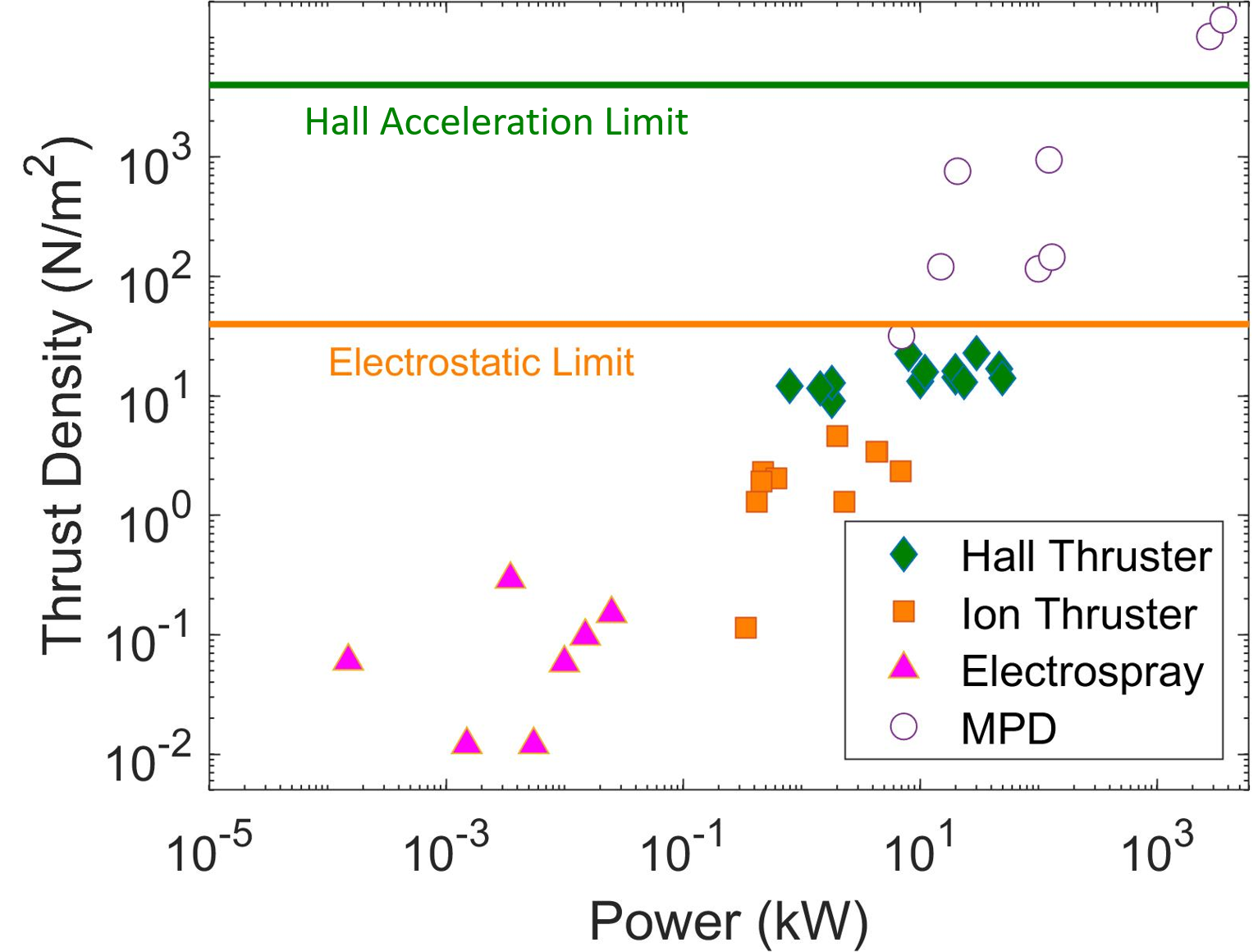}
    \caption{Measured thrust densities vs power of several forms of electric propulsion. Area is total thruster area. Thrust Density limits are shown as solid lines for Hall thrusters and electrostatic thrusters. Thrusters and references are shown in Appendix \ref{sec:AppendixMeasThrustDens}.}
    \label{fig:thrustdensities1}
\end{figure}

There remain several unanswered questions surrounding the potential limits of Hall thruster thrust density. These questions are driven by the potential benefits of operating at higher thrust densities, namely, if Hall thrusters are able to function with higher thrust density, smaller thrusters could be used on spacecraft which would allow more volume and mass for payload or fuel. It was suggested previously \cite{popov1967acceleration} that that the maximum achievable thrust density in Hall thrusters is limited by the magnetic pressure of the magnetic field in the thruster.\cite{popov1967acceleration}. This limit is typically orders of magnitude higher than the thrust density in currently operating thrusters (see Fig. \ref{fig:thrustdensities1}). Various formulations of the thrust density limits are revisited here in an attempt to reveal several constraints and dependencies on Hall thruster parameters.

This article is organized as follows. In Sec. \ref{sec:1.5_background}, an overview of previous literature on thrust density derivations is provided. In Sec. \ref{sec:2_theory}, an expression for the thrust density is derived for Hall thrusters. The thrust density is expanded into four distinct components related to distinct physical mechanisms, which are analyzed in Sec. \ref{sec:3_magpressure}: magnetic pressure and tension, Sec. \ref{sec:4_OtherTerms}: plasma resistivity, plasma pressure, and magnetic mirroring of electrons. The relative contributions of each of these components to thrust density is discussed in Sec. \ref{sec:7_contribution}, and conclusions are discussed in Sec. \ref{sec:8_conclusions}.

\section{Background}\label{sec:1.5_background}

The goal of any thruster system is to accelerate ions as the particles with a finite mass (compared to the negligible mass of electrons). At a very high level, it is useful to consider the total energy of the ions: $E_i=m_i (V_z^2+V_\theta ^2+V_r^2)/2+q \phi$. Here, $E_i$ is ion energy, $m_i$ is ion mass, $V_z$, $V_\theta$, and $V_r$ are ion velocity in the axial azimuthal and radial direction, $q$ is charge, and $\phi$ is potential energy. From this perspective, the thrust is associated with the axial ion energy due to the ion axial velocity, $V_z$. According to energy conservation, the axial ion energy can be directly converted from the electrostatic potential energy (via the axial electric field, $E_z$) as it occurs in electrostatic ion thrusters. Additional mechanisms are related to the conversion of the ion perpendicular energy into the energy of the axial motion ($\sim (V_\theta ^2+V_r^2) \rightarrow V_z^2$). Given that the static magnetic field does not change the energy of charged particles, only rotates the direction of the velocity vector, the energy must initially be gained by the electric field, particularly given ions are generally unmagnetized in Hall thrusters. The magnetic field is paramount however to create the conditions to sustain a large electric field through magnetized electrons (as in Hall thrusters), supply the perpendicular energy and convert the energy of the perpendicular motion (like rotation) into the axial energy (as in MPDs).

The deposition of energy into the ion axial motion has to be considered together with the momentum balance, in particular, with the momentum exchange between ion and electron components. In a quasineutral plasma, this momentum exchange occurs via the electric field (neglecting the ion-electron friction). Neglecting the electron inertia one can write
\begin{equation}
-en\left({\vec{E}+\vec{V}_e}\times \vec{B}\right)- \vec \nabla P_e+\vec{R}_e=0,    
\label{em}
\end{equation}
\begin{equation}
m_i \left(\vec{V}_i\cdot \vec \nabla\right) \vec{V}_i=en\left({\vec{E}+\vec{V}_i}\times \vec{B}\right)+\vec{R}_i=0.\label{im} 
   \end{equation}
   
Here $e$ is electron charge, $n$ is plasma density, $\vec{E}$ is the electric field, $\vec{V}_e$ is the electron velocity, $\vec{V}_i$ is ion velocity, $\vec{B}$ is the magnetic field, $P_e$ is the electron pressure, and $\vec{R}_e$ and $\vec{R}_i$ are the electron and ion collision momentum terms. For simplicity we omit here the ion pressure compared to the ion kinetic energy of the directed motion. For Hall thrusters, the magnetic field is often omitted but we keep it here to make connections to other systems with magnetized ions.
One can also consider the equivalent equation for the total ion+electron momentum in the form  
 
   \begin{equation}
m_i \left(\vec{V}_i\cdot \vec \nabla \right) \vec{V}_i=\vec{J}\times \vec{B}-\vec \nabla P_e+\left(\vec{R}_i+\vec{R}_e\right)=0.\label{tm} 
\end{equation}

Where $\vec J=e n (\vec V_i - \vec V_e$) is the total current. Equations (\ref{em}), (\ref{im}), and (\ref{tm}) illustrate the duality of the acceleration mechanism in Hall thruster. Equation \ref{im} shows that ions are accelerated by the electric field, hence the mechanism is electrostatic, while equation (\ref{tm}) show that the total momentum to the plasma (mostly residing in ions) is provided by the $J\times B$ force, which also applies to the magnetic system. Hence, it can be viewed as an electromagnetic acceleration given the thrust is applied to the magnets. Of course, both interpretations are correct and it is a matter of terminology \cite{Kim2015,Sasoh_Paper1}. 

Based on the definition in Eq. (\ref{tm}), the force on the plasma $\vec {F}_{EM}$ can be estimated \cite{popov1967acceleration} based on the magnetic pressure from the electromagnetic force
\begin{equation}
    \vec{F}_{EM}=\vec{J}\times \vec{B}\sim\nabla \frac{B^2}{2}.
\end{equation}
The total magnetic pressure however provides a highly overestimated limit for the thrust density. As a matter of fact, the electromagnetic force consists of the magnetic pressure and curvature forces,
\begin{equation}
    \vec{F}_{EM}=\vec{J}\times \vec{B}=-\nabla \frac{B^2}{2}+\vec{B}\cdot \vec{B},
\label{tem}
\end{equation}
which cancel each other exactly for the vacuum magnetic field. 
The total input of the momentum based on Eq. (\ref{tem})
is finite only
due to the deviation of the magnetic field from its vacuum value. This deviation is small for Hall thruster but cannot be neglected when using this expression. 

From equation (\ref{im}) and neglecting the effects of the magnetic field on ions, one can see that the momentum input on ions is due to the axial electric field. For Hall thrusters, with predominantly radial magnetic fields, the axial electric field is sustained by the azimuthal electron motion (Hall current). The axial electron stress force (pressure gradient and anisotropic stress tensor), as well as the friction force may also provide some contributions. Generating an electric field in a quasineutral plasma allows Hall thrusters to achieve thrust densities above that of gridded ion thrusters, which are limited by the Child-Langmuir law. There have been some investigations into gridded ion thruster-Hall thruster hybrids through the injection of negative ions, which purports to achieve a small increase in thrust density \cite{NegIons}.

When the axial magnetic field is included $B_z$, such as in the magnetic nozzle and cylindrical Hall thruster configurations \cite{raitses_parametric_2001}, the radial electron pressure gradient affects the azimuthal motion and thus becomes a part of the total thrust \cite{FruchtmanNozzleDerive,TakahashiNozzleDerive}. It is shown in these works that by neglecting electron-neutral collisions, which are assumed to be negligible for the magnetic nozzle thruster, the radial plasma pressure balances the azimuthal current $j_{\theta} \cross B_z$ force. This assumption cannot be made for high thrust density Hall thrusters, as electron-neutral collisions are both critical to maintaining current continuity and are shown in Section \ref{sec:3_magpressure} to be a limiting factor in the thrust density. The combined effects of the radial and axial magnetic field (in the nozzle configuration) allows the conversion of the plasma (ion) perpendicular momentum (mostly azimuthal) into the axial motion which is termed as an electromagnetic acceleration mechanism in \cite{Sasoh_Paper1,Sasoh_Paper2}. According the the terminology in \cite{Sasoh_Paper1,Sasoh_Paper2}, the configurations with diverging magnetic field (magnetic nozzle) exhibit hybrid (electrostatic and electromagnetic) acceleration mechanisms.
 
 In this manuscript, we estimate relative contributions of various contributions focusing on the geometry of the annular Hall thruster, where only the radial magnetic field is considered. The self-consistently generated magnetic field is considered in this derivation and the resulting inputs are evaluated and compared with some available experimental data. 

\section{Thrust Density Derivation}\label{sec:2_theory}

In Hall thrusters, thrust is generated by the acceleration of ions across an electric field, which is generated in the plasma by the reduced mobility of electrons in an applied magnetic field. The thrust density can be related to this electric field by integrating the product of the charge $e$, the plasma density $n_e$ and the axial component of the electric field $E_z$ over the axial domain $z_0$ to $z_f$, and some front-facing area $A$:

\begin{equation}
    T = \int^{A}{\int_{z_o}^{z_f}{e n_e E_z dz}  dA}\,.
\end{equation}

The thrust density for the thruster is then simply
\begin{equation}
    \mathcal T = \frac{T}{A} = \int_{z_o}^{z_f}{e n_e E_z dz}\,.
\end{equation}

The electric field $E$ can be solved by utilizing the steady-state electron fluid momentum balance equations. This is done under the assumption of electron momentum continuity, steady-state operation of the thruster, and negligible contribution of electron-electron and electron-ion collisions.

\begin{equation}
    0 = -e n_e (\vec E + \vec u \times \vec B) - m_e n_e \nu_{en} \vec u - \vec \nabla \cdot \bar{\bar P}\,,
\end{equation}

\begin{equation}
    \vec E = -(\vec u \times \vec B) - \frac{m_e \nu_{en}}{e} \vec u - \frac{1}{e n_e}\vec \nabla \cdot \bar{\bar P}\,,
\end{equation}
where $u$ is the electron fluid velocity, $B$ is the magnetic field, $m_e$ is the mass of an electron, $\nu_{en}$ is the electron-neutral collision frequency, and $\bar{\bar P}$ is the electron plasma pressure tensor. If we at first define our coordinate system by the parallel and perpendicular components of the magnetic field,\cite{moore_magnetic_1988} we can expand the pressure term by the following:

\begin{equation}
\begin{aligned}
    E_{\parallel} &= -(\vec u \times \vec B)_{\parallel} - \frac{m_e \nu_{en}}{e} \vec u_{\parallel} - \frac{1}{e n_e}(\vec \nabla \cdot \bar{\bar P})_{\parallel} \\
    &= 0 - \frac{m_e \nu_{en}}{e} \vec u_{\parallel} - \frac{1}{e n_e}\left(\nabla_{\parallel}P_{\parallel} - \left(\frac{P_{\parallel}-P_{\perp}}{B}\right)\nabla_{\parallel}B   \right)\,,\\
\end{aligned}
\end{equation}

\begin{equation}
\begin{aligned}
    E_{\perp} &= -(\vec u \times \vec B)_{\perp} - \frac{m_e \nu_{en}}{e} \vec u_{\perp} - \frac{1}{e n_e}(\vec \nabla \cdot \bar{\bar P})_{\perp} \\
    &= -(\vec u \times \vec B)_{\perp} - \frac{m_e \nu_{en}}{e} \vec u_{\perp} - \frac{1}{e n_e}\left(\nabla_{\perp}P_{\perp}\right)\,.\\
\end{aligned}
\end{equation}

To return to cylindrical coordinates for the purposes of calculating $E_z$ for a thruster, we must take into account some angle $\alpha$ as shown in Fig. \ref{fig:EzvsB}. The z component is simply the parallel and perpendicular of the previous equations scaled by the cosine and sine of $\alpha$ respectively:

\begin{figure}
    \centering
    \includegraphics{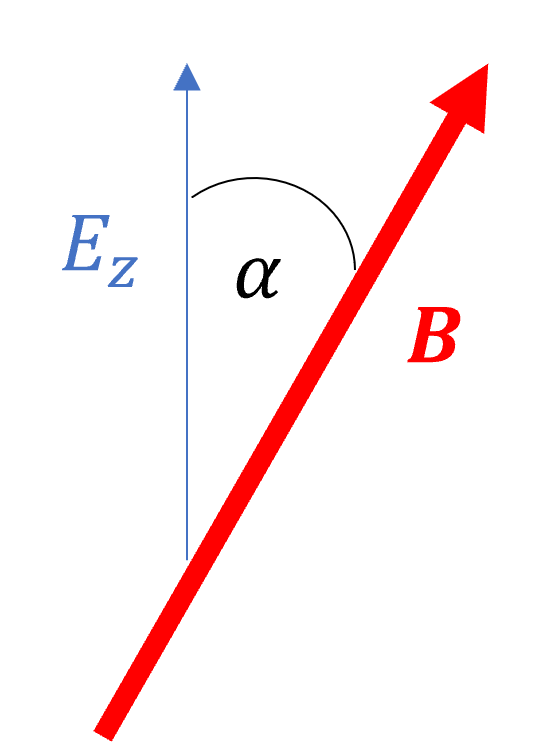}
    \caption{Angle between magnetic field and axial electric field}
    \label{fig:EzvsB}
\end{figure}

\begin{equation}
\begin{aligned}
        E_z &= -(\vec u \times \vec B)_z - \frac{m_e \nu_{en}}{e} \vec u_z - \frac{1}{e n_e}\left(\nabla_{\parallel}P_{\parallel} - \left(\frac{P_{\parallel}-P_{\perp}}{B}\right)\nabla_{\parallel}B\right) \cos{\alpha} \\
        &- \frac{1}{e n_e}\left(\nabla_{\perp}P_{\perp}\right)\sin{\alpha} \\
        &= -(\vec u \times \vec B)_z - \frac{m_e \nu_{en}}{e} \vec u_z - \frac{1}{e n_e}\left(\nabla_z (P_{\parallel}\cos^2{\alpha} + P_{\perp}\sin^2{\alpha})\right) \\
        &- \frac{\cos^2{\alpha}}{e n_e} \left(\frac{P_{\perp}-P_{\parallel}}{B}\right)\nabla_z B\,.
\end{aligned}
\end{equation}

We will then convert the electron fluid velocities into current by the relation $\vec j = - e n_e \vec u$:

\begin{equation}\label{eq:Ez_components}
\begin{aligned}
    E_z &= \frac{1}{e n_e}(\vec j \times \vec B)_z  + \frac{m_e \nu_{en}}{e^2 n_e} \vec j_z - \frac{1}{e n_e}\left(\nabla_z (P_{\parallel}\cos^2{\alpha} + P_{\perp}\sin^2{\alpha})\right) \\
    &- \frac{\cos^2{\alpha}}{e n_e} \left(\frac{P_{\perp}-P_{\parallel}}{B}\right)\nabla_z B\,.
    \end{aligned}
\end{equation}

The electric field is then due to the sum of four components in Eq. (\ref{eq:Ez_components}): magnetic pressure/tension, axial resistivity pressure, plasma pressure, and magnetic mirror pressure. We will now solve each individually:

\begin{equation}
    E_z = E_{z,1} + E_{z,2} + E_{z,3} + E_{z,4}\,.
\end{equation}

\section{Magnetic Pressure and Tension}\label{sec:3_magpressure}
The first component, which we will soon find is related to the magnetic pressure and tension, is
\begin{equation}\label{eq:Ez1begin}
    E_{z,1} = \frac{1}{e n_e}(\vec j \times \vec B)_z\,.
\end{equation}

The electron current in the plasma is not known, however we are able to determine the maximum current possible before the induced magnetic field overwhelms the externally applied magnetic field. The current can be related to the curl of the magnetic field through Maxwell's equations:

\begin{equation} \label{eq:maxwells}
    \vec j = \frac{1}{\mu_0} \vec \nabla \times \vec B\,,
\end{equation}
where $\mu_0$ is the magnetic permeability of free space. Substituting Eq. (\ref{eq:maxwells}) into Eq. (\ref{eq:Ez1begin}):

\begin{equation}\label{eq:thrustdensmagelec}
    \begin{aligned}
        E_{z,1} &= \frac{1}{e n_e}(\vec j \times \vec B)_z \\
        &=\frac{1}{e n_e}\left(\frac{1}{\mu_0} (\vec \nabla \times \vec B ) \times \vec B\right)_z \\
        &= \frac{1}{e n_e}\left( -\frac{\vec \nabla (B^2)}{2 \mu_0} + (\vec B \cdot \vec \nabla)\vec B \right)_z \\
        &= -\frac{1}{e n_e}\left(\frac{\vec \nabla_{\perp}(B^2)}{2 \mu_0} - \frac{B^2}{\mu_0}\frac{\hat R_B}{R_B}  \right)_z\,,\\
    \end{aligned}
\end{equation}
where the first term on the right hand side (RHS) is the magnetic pressure in the system and the second term on the RHS is the magnetic tension, and $\hat R_B$ is the outward pointing vector of the radius of curvature of the magnetic field. For straight radial magnetic fields with no curvature, both the magnetic pressure and tension are zero, which is often the idealized assumption for the field formed in an annular Hall thruster channel. For magnetic fields with finite curvature, which is a typical case in Hall thrusters, if there is no current-induced magnetic field the pressure and tension balance each other such that they sum to zero. This continues to be the case for thrusters with strong axial fields, such as cylindrical Hall thrusters,\cite{raitses_parametric_2001} although the magnitude of both the magnetic pressure and tension terms tend to be higher due to the stronger gradients and curvature.

\begin{figure}
    \centering
    \includegraphics[width=0.75\textwidth]{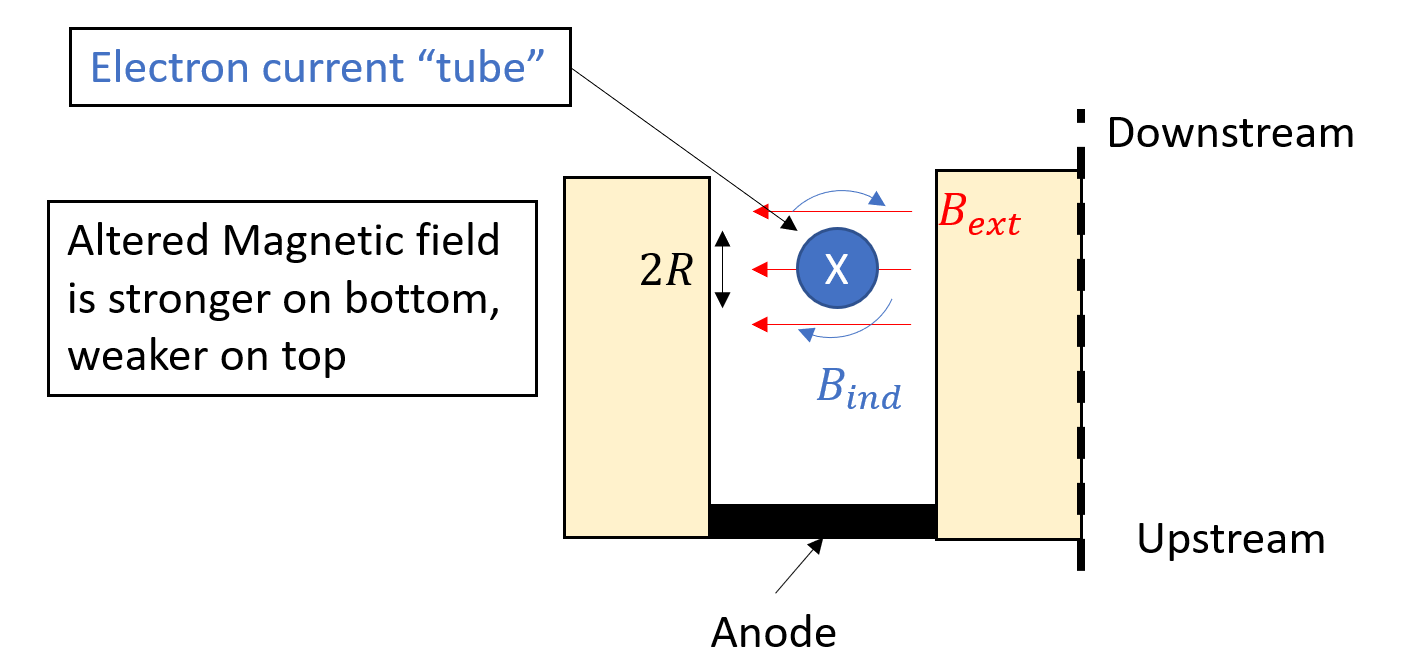}
    \caption{Sketch of an annular Hall thruster with azimuthal current with induced magnetic field}
    \label{fig:annularsketch}
\end{figure}

The limit of the magnetic pressure and tension term can be derived for the case of a simplified annular Hall thruster configuration. Consider an annular Hall thruster with a magnetic field which only has a radial component, such as shown in Figure \ref{fig:annularsketch}. In this scenario, the resulting electric field is entirely in the axial direction, and so the plasma current forms in the ExB direction. The ExB plasma current travels azimuthally around the thruster channel in a closed loop, inducing a magnetic field of some magnitude $B_{ind}$. The induced magnetic field vectors and the externally applied magnetic field vectors can be added to determine the total magnetic field at any point. The magnitude of the induced magnetic field within the azimuthal current density $j_{\theta}$ and area $A$ can be written, by the Biot–Savart law, as:

\begin{equation}
    B_{ind}(z) = \frac{\mu_0 j_{\theta} A}{2 \pi \abs{z}} = \frac{\mu_0 j_{\theta} \pi \abs{z}^2}{2 \pi \abs{z}} = \frac{\mu_0 j_{\theta} \abs{z}}{2}\,,
\end{equation}
where $z$ is the axial distance from the center of the current tube. Note that the absolute value is used here as the axial term is used in substitution of the conventional radial term of the Bio-Savart law. If we consider the center of the current loop to be our axial zero, we can define the outer radius of this current loop to be $R$. Given the orientation of the magnetic fields shown in Fig. \ref{fig:annularsketch}, the total magnetic field in the downstream side and the upstream side of the current loop are simply:
\begin{equation}
\begin{aligned}
    B(z : -R \rightarrow 0) &= B_{ext}+B_{ind}(z)\,, \\ 
    B(z : 0 \rightarrow +R) &= B_{ext}-B_{ind}(z)\,.
\end{aligned}
\end{equation}
At the edges of the current tube $z=\pm R$, the magnitude of the induced magnetic field reaches a maximum and can be defined as $B_{ind}(z=\pm R) = B_{IND}$. 

The component of thrust density due to magnetic pressure in Eq. (\ref{eq:thrustdensmagelec}) can then be written in terms of $B_{IND}$ and $B_{ext}$:

\begin{equation}\label{eq:magpressurederive}
\begin{aligned}
    \mathcal T_1 &= \int_{z_o}^{z_f}{e n_e E_{z,1} dz} \\
    &= \int_{z_o}^{z_f}{(\vec j \times \vec B)_z dz} \\
    &= -\int_{z_o}^{z_f}{\left(\frac{\vec \nabla_{\perp}(B^2)}{2 \mu_0} - \frac{B^2}{\mu_0}\frac{\hat R_B}{R_B}  \right)_z dz} \\ 
    &= -\int_{z_o}^{z_f}{\left(\frac{1}{2 \mu_0}\frac{d(B^2)}{dz} + \frac{B^2}{\mu_0}\frac{1}{R_B}  \right) dz}\\
    &= -\int_{B(-R)}^{B(R)}{ \frac{d(B^2)}{2\mu_0}} - \int_{z_o}^{z_f}{\left(\frac{B^2}{\mu_0}\frac{1}{R_B}  \right) dz} \\
    &= \frac{1}{2 \mu_0} \left( (B_{ext}+B_{IND})^2 - (B_{ext}-B_{IND})^2\right) - \int_{z_o}^{z_f}{\left(\frac{B^2}{\mu_0}\frac{1}{R_B}  \right) dz} \\
    &= \frac{1}{2 \mu_0} \left( (B_{ext}^2+B_{IND}^2+2B_{IND} B_{ext}) - (B_{ext}^2+B_{IND}^2-2B_{IND} B_{ext})\right) \\
    &- \int_{z_o}^{z_f}{\left(\frac{B^2}{\mu_0}\frac{1}{R_B}  \right) dz} \\
    &= \frac{1}{2 \mu_0} \left( 4 B_{IND} B_{ext} \right) - \int_{z_o}^{z_f}{\left(\frac{B^2}{\mu_0}\frac{1}{R_B}  \right) dz}\,.
    \end{aligned}
\end{equation}

Thus in this formulation of an idealized Hall thruster with a simple current loop, the magnetic pressure term scales with both the applied magnetic field $B_{ext}$ and with the induced magnetic field $B_{IND}$ due to the Hall current. The limits of the thrust density due to magnetic pressure are accordingly tied to the limits of these two quantities. 

The magnetic tension term requires significantly more simplification due to the radius of curvature term. The radius of curvature $R_B$ of the total magnetic field can be written as a function of both the magnitude and radius of curvature of the external and induced fields, which is derived in Appendix \ref{sec:AppendixCurve}. It is noted that the radius of curvature of the induced field is simply the distance from the center of the current tube ($R_{B,ind} = z <= R$) and that the radius of curvature of the externally applied field in our ideal model is much larger than the radius of the electron current and correspondingly the radius of curvature of the induced field ($R_{ext}>>R$). A generalized form of thrust density due to magnetic curvature without the latter assumption is found in Appendix \ref{sec:AppendixCurve}. This curvature will be slightly different for the top of the current tube as compared to the bottom, given the induced fields are opposing the applied field on the top and in the same direction on the bottom:

\begin{equation}\label{eq:Rtopsimp}
    \frac{1}{R_{top}} = \frac{\frac{B^2_{ext}}{R_{B,ext}} + \frac{B^2_{ind}}{R_{B,ind}} - B_{ind}B_{ext}\left(\frac{1}{R_{B,ind}}\right)}{(B_{ext}-B_{ind})^2}\,,
\end{equation}

\begin{equation}\label{eq:Rbotsimp}
    \frac{1}{R_{bot}} = \frac{\frac{B^2_{ext}}{R_{B,ext}} + \frac{B^2_{ind}}{R_{B,ind}} + B_{ind}B_{ext}\left(\frac{1}{R_{B,ind}}\right)}{(B_{ext}+B_{ind})^2}\,.
\end{equation}

We can substitute the radii of curvature of Eq. (\ref{eq:Rtopsimp}) and Eq. (\ref{eq:Rbotsimp}) into Eq. (\ref{eq:magpressurederive}) to integrate the magnetic tension term over the radius of the current tube $R$, noting that the orientation of the radius of curvature is negative from $z_0$ to $0$ and positive from $0$ to $z_f$ (Fig. \ref{fig:annularsketch}) and the radius of curvature of the induced field is simply the distance from the center of the current tube $z$:

\begin{equation}\label{eq:longmagpressderive}
\begin{aligned}
    \mathcal T_1 &= \frac{4 B_{IND} B_{ext}}{2 \mu_0} + \int_{0}^{R}{\left(\frac{(B_{ext} + B_{ind})^2}{\mu_0 R_B}  \right) dz} - \int_{0}^{R}{\left(\frac{(B_{ext} - B_{ind})^2}{\mu_0 R_B}  \right) dz} \\
    &= \frac{4 B_{IND} B_{ext}}{2 \mu_0} + \frac{1}{\mu_0}\int_{0}^{R}{\left(\frac{B^2_{ext}}{R_{B,ext}} + \frac{B^2_{ind}}{R_{B,ind}} + B_{ind}B_{ext}\left(\frac{1}{R_{B,ind}}\right) \right)dz} \\
    &- \frac{1}{\mu_0}\int_{0}^{R}{\left(\frac{B^2_{ext}}{R_{B,ext}} + \frac{B^2_{ind}}{R_{B,ind}} - B_{ind}B_{ext}\left(\frac{1}{R_{B,ind}}\right) \right)dz} \\
    &= \frac{4 B_{IND} B_{ext}}{2 \mu_0} + \frac{2}{\mu_0}\int_{0}^{R}{\left(B_{ind}B_{ext}\left(\frac{1}{z}\right) \right)dz} \\
    &= \frac{4 B_{IND} B_{ext}}{2 \mu_0} + \frac{2}{\mu_0}\int_{0}^{R}{\left(\frac{\mu_0 j z}{2} B_{ext}\left(\frac{1}{z}\right) \right)dz} \\
    &= \frac{4 B_{IND} B_{ext}}{2 \mu_0} + \frac{2}{\mu_0}\frac{\mu_0 j}{2}B_{ext}\int_{0}^{R}{dz} \\
    &= \frac{4 B_{IND} B_{ext}}{2 \mu_0} + \frac{2}{\mu_0}\frac{\mu_0 j}{2}B_{ext}\left(R\right) \\
    &= \frac{4 B_{IND} B_{ext}}{2 \mu_0} + \frac{4 B_{IND} B_{ext}}{2 \mu_0}\,. \\
    \end{aligned}
\end{equation}

Thus the total thrust density due to the magnetic field is the sum of the magnetic pressure and the magnetic tension, which have equivalent magnitudes for the idealized annular Hall thruster. Essentially the total magnetic field increases on the upstream side of the current tube and decreases on the downstream side, which creates an axially positive magnetic pressure. The current also induces a curl in the local magnetic field, and the magnetic tension creates a pressure pointing towards the center of the current to straighten this magnetic field. Given that the tension scales by the square of the magnetic field and the magnetic field is higher on the upstream side than the downstream, the net tension points in the positive axial direction and also contributes to thrust. Note that the relative scaling between the magnetic pressure and curvature terms change with the geometry of the magnetic field.

The limit of this component of thrust density can be calculated by considering the limits of the applied and induced magnetic field in the system. The applied field is limited by the saturation of the magnetic field in the high-permeability material of the magnetic circuit. While this is dependent on the actual geometry of the magnetic circuit, a rough estimate using commonly low-carbon steel provides a limit of $B_{ext} = 1$ kG in Hall thruster channels. As the induced field increases, it nullifies the externally applied field in the downstream portion of the current tube. This will reduce the axial confinement of the electrons and eventually transition the operation to that of an MPD thruster rather than a Hall thruster due to the induced azimuthal magnetic fields. This hybrid configuration due to higher mass flow and plasma density has been explored recently in Ref. \cite{Sasoh_Paper1,Sasoh_Paper2}. For a Hall thruster, the current is primarily in the azimuthal direction which is possible due to the much larger electron gyrofrequency $\omega_{ge}$ compared to electron cross-field collision frequency $\nu$, the ratio of which is often called the Hall parameter $\Omega_H$. The minimum value of this Hall parameter $\Omega_{H,min}$ is dependent on the geometry of the thruster, and is derived in Appendix \ref{sec:AppendixHall}.

\begin{equation}\label{eq:hallparam}
    \Omega_H = \frac{\omega_{ge}}{\nu} > \Omega_{H,min}\,,
\end{equation}
\begin{equation}
    \omega_{ge} = \frac{e B}{m_e}\,.
\end{equation}

The collision frequency of electrons is the sum of electron-neutral collision frequency ($\nu_{en}$) and an anomalous collision frequency ($\nu_{anom}$) due to Bohm diffusion, which is typically ascribed to plasma turbulence \cite{boeuf_tutorial_2017}. The anomalous collision frequency scales by some Bohm constant $\kappa_B$ and the electron gyrofrequency:

\begin{equation}
    \nu = \nu_{en} + \nu_{anom} = \nu_{en} + \frac{\kappa_B \omega_{ge}}{16}\,.
\end{equation}

The value of $\kappa_B$ has been experimentally shown to be about 0.1 in annular Hall thruster plasmas \cite{boeuf_tutorial_2017}, however has shown to be higher in other thruster types such as cylindrical Hall thrusters\cite{smirnov_electron_2006}. If we use the inequality in Eq. (\ref{eq:hallparam}) and take into account the induced magnetic field, we can find a limit on the induced field and consequently a limit on the thrust density. The maximum induced field a Hall thruster can contain without losing confinement of electrons, and while retaining Hall-thruster operation, can be written as:

\begin{equation}
\begin{aligned}
    \frac{\omega_{ge}}{\nu} &> \Omega_{H,min}\,, \\
    \frac{\omega_{ge}}{\nu_{en} + \frac{\kappa_B \omega_{ge}}{16}} &> \Omega_{H,min}\,, \\
    \frac{\frac{e (B_{ext}-B_{IND})}{m_e}}{\nu_{en} + \frac{\kappa_B e (B_{ext}-B_{IND})}{16}} &> \Omega_{H,min}\,, \\
    B_{IND} &< B_{ext} - \frac{\Omega_{H,min}\nu_{en}m_e}{e\left(1-\Omega_{H,min}\frac{\kappa_B}{16}\right)}\,.
\end{aligned}
\end{equation}

\begin{figure}
    \centering
    \includegraphics[width=0.75\textwidth]{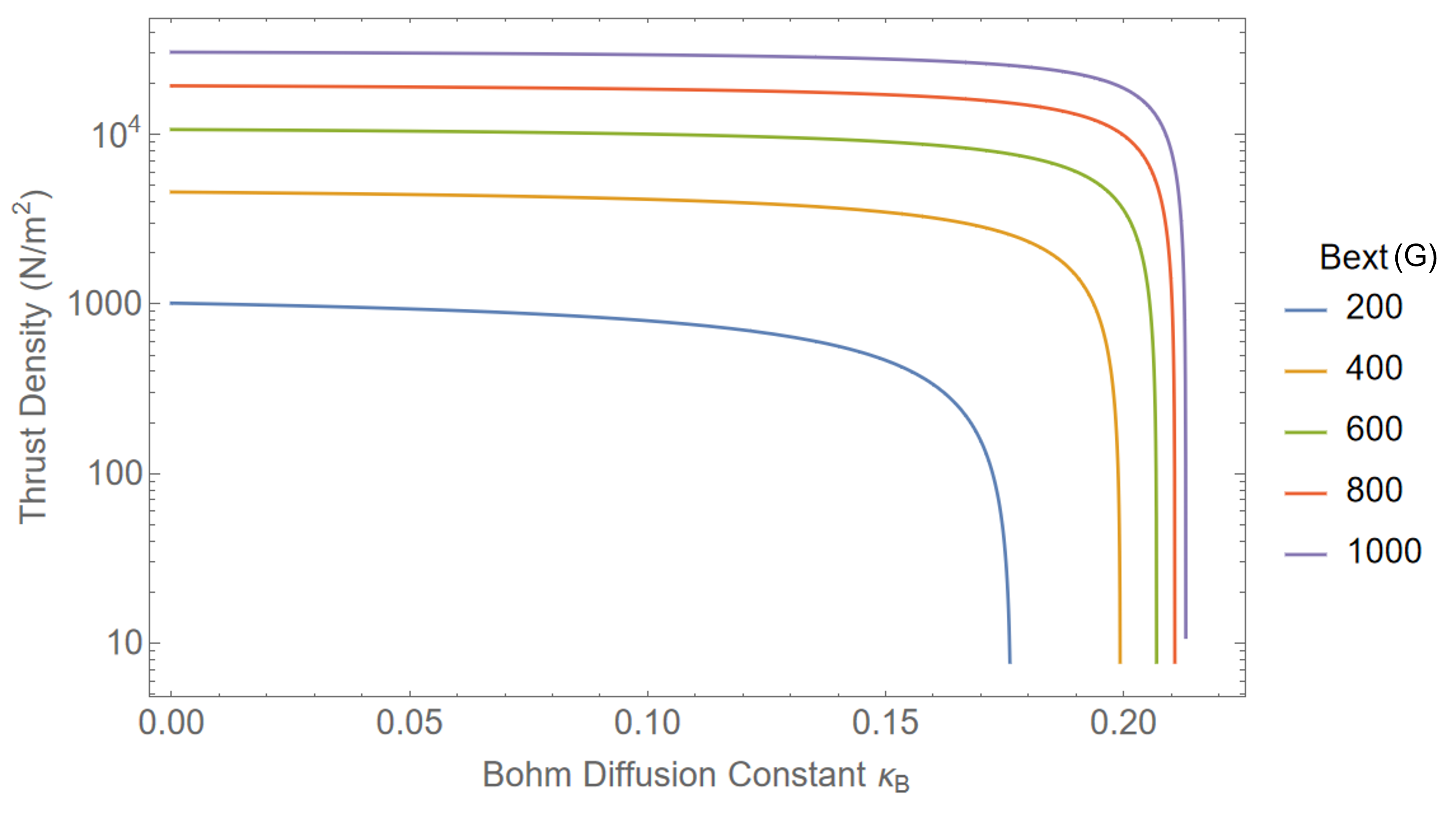}
    \caption{Maximum Thrust Density due to Magnetic Pressure and Tension for varying externally applied magnetic field (in Gauss) and Bohm diffusion Constant $\kappa_B$, where $\Omega_{H,min} = 72$ and $\nu_{en} = 10^7$ Hz}
    \label{fig:thrustdensmagpressure}
\end{figure}

The induced magnetic field must be smaller than the externally applied magnetic field by a factor that scales with the Bohm diffusion. For  $\Omega_{H,min} \sim 100$ and $\kappa_B \sim 0.1$, this results in $B_{IND} \sim B_{ext}-153 G$, with a corresponding thrust density of 880 N/m$^2$. Plots of the maximum thrust density while retaining Hall-thruster operation with varying magnetic fields and Bohm diffusion constant $\kappa_B$ are shown in Figure \ref{fig:thrustdensmagpressure}, where it can be observed that $\kappa_B$ must be small to maintain operation of the thruster, and the thrust density collapses if the Bohm diffusion is too high.

In general, $\mathcal T_1$ is the largest component of the thrust density, however the other three components will be derived for completeness. Relative values of each component will be shown in Sec. \ref{sec:7_contribution}. 

\section{Other Contributions}\label{sec:4_OtherTerms}
\subsection{Axial Plasma Resistivity}
The second component is related to the plasma resistivity for the axial electron current.
\begin{equation}
    E_{z,2} = \frac{m_e \nu_{en}}{e^2 n_e} \vec j_z\,,
\end{equation}

\begin{equation}
    \mathcal T_2 = \int_{z_o}^{z_f}{e n_e E_{z,2} dz} = \frac{m_e}{e}\int_{z_o}^{z_f}{\nu_{en}j_z dz}\,.
\end{equation}

This value is usually very small due to the low values of electron-neutral collision frequency. Utilizing a thruster in this manner is highly inefficient due to the energy losses to the neutrals, however it does represent some contribution to the thrust density. Thrust due to this plasma resistivity may also became appreciable in poor vacuum conditions as the neutral density increases. Higher thrust has been observed in Hall thrusters with higher neutral background density \cite{FruchtmanCollisional}, however this may also be due to charge-exchange collisions. It should be noted that as plasma density increases, the electron-ion and ion-neutral collisions become appreciable. These are of particular importance to the operation of MPD thrusters, and while this effect is not investigated here, future work may analyze the thruster operation of a Hall thruster-MPD hybrid, which may occur at high thrust densities as was explored in recent work \cite{Sasoh_Paper1,Sasoh_Paper2}.

\subsection{Plasma Pressure}

The third component due to plasma pressure is:
\begin{equation}
    E_{z,3} = - \frac{1}{e n_e}\left(\nabla_z (P_{\parallel}\cos^2{\alpha} + P_{\perp}\sin^2{\alpha})\right)\,,
\end{equation}
where it should be noted that when the electron energy is isotropic such that $P_{\parallel} = P_{\perp}$, this term reduces to the more familiar form:

\begin{equation}
    E_{z,3} = - \frac{1}{e n_e} \frac{dP}{dz}\,.
\end{equation}

The integral for plasma pressure thrust density is dependent on the angle of the magnetic field over the axial span when considering the anisotropic case, and so for a simplified case of isotropic electron energy, this thrust density component can be derived to be:

\begin{equation}
    \mathcal T_3 = \int_{z_o}^{z_f}{e n_e E_{z,3} dz} = -\int_{P_{max}}^{0}{dP} = P_{max}\,,
\end{equation}
where the minimum plasma pressure is assumed to be zero, as the thruster is operating in vacuum.

\subsection{Magnetic Mirror}\label{sec:6_mirror}
The fourth component is the pressure to the magnetic mirror, which arises when a strong anisotropic electron energy is present in the plasma:
\begin{equation}
    E_{z,4} = - \frac{\cos^2{\alpha}}{e n_e} \left(\frac{P_{\perp}-P_{\parallel}}{B}\right)\nabla_z B\,.
\end{equation}

Much like the plasma pressure term, this is dependent on the angle of the magnetic field, and in general will primarily exist in the center of thrusters where the magnetic field is primarily axial and diverging. Wall-less Hall thrusters and cylindrical Hall thrusters have such conditions, however the anisotropy of the electron energy in the central region is under-investigated. Assuming the angle of the magnetic field over the axial path is relatively constant, and assuming the plasma pressure does not change over the integration path, the thrust density can be derived to be:

\begin{equation}
\begin{aligned}
    \mathcal T_4 &= \int_{z_o}^{z_f}{e n_e E_{z,4} dz} \\
    &= -\int_{B_{IND}}^{B_{min}}{\cos^2{\alpha} \left(\frac{P_{\perp}-P_{\parallel}}{B}\right) dB} \\
    &= \cos^2{\alpha} \left(P_{\perp}-P_{\parallel}\right) \ln{\left(\frac{B_{max}}{B_{min}}\right)}\,.
\end{aligned}
\end{equation}

This is quite similar in form to the thrust derived in Ref. \cite{FruchtmanNozzleDerive} for magnetic nozzle thrusters, however in that case the plasma pressure was assumed to be isotropic.

\section{Relative Contributions}\label{sec:7_contribution}
The relative magnitude of each of the four components of thrust density derived for Hall thrusters can be estimated using typical operating conditions using the assumptions in Table \ref{table:thrustdensassumptions}, which were obtained from experimental measurements in a 2 kW Hall thruster \cite{staack_shielded_2004}. The magnetic mirror term was assumed to be zero given the radially dominant fields ($\alpha \sim 0$) and due to the absence of electron temperature anisotropy measurements.

\begin{table}[ht]
\caption{Assumptions of the thrust density magnitudes}
\centering
\begin{tabular}{|p{0.2\linewidth} | p{0.2\linewidth} |}
 \hline
Element & Value \\ [0.5ex] 
 \hline\hline
 $n_e$ & $5 \cdot 10^{17}$ m$^3$ \\ 
 \hline
 $T_{e}$ & 30 eV \\ 
 \hline
 $j_z$ & 0.04 A/cm$^2$ \\ 
 \hline
 $\nu_{en}$ & 10$^7$ Hz \\ 
 \hline
  $z_f-z_0$ & $0.5$ cm \\ 
 \hline
   $V_D$ & $250$ V \\ 
 \hline
\end{tabular}
\label{table:thrustdensassumptions}
\end{table}

While the value of the thrust density due to axial resistivity, plasma pressure, and the mirror force are relatively straightforward, the magnetic pressure term requires some knowledge of the induced magnetic field, which is not inherently obvious. However an estimate of this induced field can be derived from the applied voltage and the plasma density by the following:

\begin{equation}
    j_{\theta} = \frac{e n_e E}{B_{ext}} = \frac{e n_e}{B_{ext}}\frac{V_D}{2R}\,,
\end{equation}

\begin{equation}\label{eq:BindToBext}
    B_{IND} = \frac{\mu_0 j_{\theta} R}{2}=\frac{\mu_0 R}{2}\frac{e n_e}{B_{ext}}\frac{V_D}{2R}=\frac{ e n_e \mu_0 V_D}{4 B_{ext}}\,.
\end{equation}

Here we have assumed the potential drop occurs along the length of the electron tube of diameter $2R$. By substituting $B_{IND}$ into Eq. (\ref{eq:longmagpressderive}), one can simplify the expression to determine the estimated value of the magnetic pressure and tension thrust density given operating parameters:

\begin{equation}
    \begin{aligned}
        \mathcal T_1 = e n_e V_D\,,
    \end{aligned}
\end{equation}
The values of thrust density for each component is tabulated in Table \ref{table:thrustdensmag} given the typical operating condition in Table \ref{table:thrustdensassumptions}, where one may observe that the Magnetic pressure term dominates conventional operation.

\begin{table}[ht]
\caption{Relative Magnitude of Thrust Density Components}
\centering
\begin{tabular}{|p{0.3\linewidth} |p{0.2\linewidth} | p{0.2\linewidth} |}
 \hline
Term & Element & Value \\ [0.5ex] 
 \hline\hline
 Magnetic Pressure & $\mathcal T_1$ & $\sim$ 20 N/m$^2$ \\ 
 \hline
 Plasma Resistivity & $\mathcal T_2$ & $\sim$ 0.001 N/m$^2$ \\ 
 \hline
 Plasma Pressure & $\mathcal T_3$ & $\sim$ 2.4 N/m$^2$ \\ 
  \hline
 Magnetic Mirror & $\mathcal T_4$ & $\sim$ 0 N/m$^2$ \\ 
 \hline
 Total & $\mathcal T$ & $\sim$ 22 N/m$^2$ \\ 
   \hline
\end{tabular}
\label{table:thrustdensmag}
\end{table}

Given the simple estimation of Eq. \ref{eq:BindToBext}, it may appear that the thrust density is only dependent on the applied voltage and is not related to the either the external or induced magnetic field. This is not the case, as the accelerating electric field ultimately is due to the axial reduction of electron mobility due to the magnetic field. The role of $B_{ext}$ in the thrust generation is then to maintain this electron confinement. $B_{ind}$ on the other hand, is a proxy for the plasma density in this confined region, and both directly scale with the thrust density as there are more ions to accelerate, but this higher plasma density also decreases the total magnetic field until confinement no longer occurs. $B_{ext}$ is limited by the magnetic circuit of the thruster. The high magnetic permeability materials of modern thrusters typically saturate with applied magnetic fields of around 20 kG. Given the drop in magnetic flux through the air gap of the discharge channel, the resulting $B_{ext}$ along the channel centerline is typically below 1 kG. $B_{ind}$ is largely correlated with the mass flow through the thruster, as the plasma density scales with the mass flow. Limitations on increasing the plasma density and induced field are usually due to thermal management in Hall thrusters, as the heat from the electrons colliding with the channel wall flows to the magnetic circuit and thruster coils/permanent magnet, the latter of which typically have maximum temperatures of $\sim 400 ^{\circ}$C. This appears to be the most likely area for future improvements in thrust density, as better thermal designs may allow much higher mass flows and higher thrust densities accordingly.

\subsection{Role of Plume Divergence}
It should be noted that the thrust density is dependent on the axial component of the electric field, as the radial terms sum to zero due to the axial symmetry of these devices. The radial component of the electric field is ultimately a loss mechanism, as energy is invested into radial acceleration of ions that does not contribute to thrust. The divergence of the plume, which is commonly referred to by an angle at which some fraction of the ion population is accelerated, is a measure of the thrust which is lost. If the plume divergence is defined by the momentum weighted angle at which 50\% of ions are accelerated ($\theta_{mom}$), and we assume the ion velocity is not dependent on the acceleration angle, we can define the proportion of thrust density lost to plume divergence. This would be:
\begin{equation}
    \cos{(\theta_{mom})} = \frac{2 \pi \int_{0}^{\pi/2}{j_i \sin{\theta}\cos{\theta}d\theta}}{2 \pi \int_{0}^{\pi/2}{j_i \sin{\theta}d\theta}}\,,
\end{equation}

\begin{equation}
    \mathcal T_{divergence} = \cos{(\theta_{mom})}\mathcal T_{ideal}\,.
\end{equation}
Where $j_i$ is the ion current, $\theta$ is the plume angle, $\mathcal T_{ideal}$ is the ideal thrust density assuming no divergence, and $\mathcal T_{divergence}$ is the thrust density taking into account this divergence. This factor directly lowers the thrust density of the thruster.

\section{Conclusion}\label{sec:8_conclusions}

The formulation of thrust density provides some insight into the potential limits of Hall thruster operation. In general, the magnetic pressure and tension term provides the majority of thrust density in annular Hall thrusters, however the magnetic mirror term may provide appreciable magnitudes for thrusters with diverging magnetic fields as shown in recent experiments.\cite{simmonds_ion_2021}. The plasma resistivity and pressure terms, while important to the operation of the thruster itself, provide relatively negligible thrust. The limits of thrust density show magnitudes well above that calculated by the conventional operation as defined in Table \ref{table:thrustdensassumptions}. 

When comparing the calculated thrust density using the derived expressions in Table \ref{table:thrustdensmag} to the measured thrust density in many Hall thrusters, as shown in Figure \ref{fig:thrustdensities1}, a clear linear trend is apparent between measured thrust density and power. This is curious given that this trend occurs over all forms of electric propulsion, despite the vastly different acceleration mechanisms. Hall thrusters appear an outlier in this regard, as rather than following this trend they tend to operate with relatively constant thrust densities of $\sim 10$ N/m$^2$ over a large range of power levels. Given that this is several orders of magnitude below their potential thrust density limit of $\sim 10^3$ N/m$^2$, there appears to be great potential for improvement of these devices through miniaturization or increasing throughput of conventional sized devices. 

It is notable that no Hall thruster has been operated at such higher thrust densities as of writing this article, largely due to material limitations, and the behavior of the plasma and the role of instabilities at these levels is not well understood. These instabilities have not hampered Hall thruster operation yet, however given the relation between the anomalous collision frequency and the maximum thrust density in Fig. \ref{fig:thrustdensmagpressure}, minimizing the anomalous collision frequency of electrons may soon be a critical challenge to operating at higher thrust density levels.  The path forward appears to be one part an improvement in the engineering of the devices and another part an improvement in the understanding of the physics. The heat management of the devices needs to significantly improve to maintain the magnetic circuit at these high plasma densities, and the anomalous electron transport in the plasma needs to be reduced, likely by suppression of the plasma instabilities. Yet, despite the potential challenges that these may represent, it is clear we are operating these thrusters well below their potential capability.

\section*{Declarations}

\subsection*{Acknowledgments}
The authors would like to acknowledge Igor Kaganovich and Nirbhav Chopra for fruitful discussions.

\subsection*{Competing interests}
The authors have no relevant financial or non-financial interests to disclose.

\subsection*{Author's contributions}
All authors contributed to the study conception and design. The first draft of the manuscript was written by Jacob Simmonds and all authors commented on previous versions of the manuscript. All authors read and approved the final manuscript

\subsection*{Funding}
This work was supported by the AFOSR under contract FA9550-17-1-0010 and the US Department of Energy under contract DE-AC02-09CH11466.

\subsection*{Availability of data and materials}
All data generated or analysed during this study are included in this published article.

\begin{appendices}

\section{Compiled List of Thrust Densities}\label{sec:AppendixMeasThrustDens}

Thrust density for the thrusters in Figure \ref{fig:thrustdensities1} is defined as the measured thrust in literature divided by the total front-facing area, as compiled in Table \ref{table:thrustdensreference}.

\begin{table}\label{table:thrustdensreference}
\caption{Compiled list of thrust densities and power levels of several forms of electric propulsion in descending order of measured thrust density.}
\centering
\begin{tabular}{ |c|c|c|c|c| } 
  \hline
 MPD Thruster & Organization & Thrust Density & Power & Reference \\ 
   &  & (N/m$^2$) & (kW) & \\ 
 \hline
 MY-III & Osaka University & 14,012 & 3,630 & \cite{MPDCollection} \\
 MultiMegawatt MPD & Princeton University & 10,186 & 2,800 & \cite{Multimegawatt_MPD} \\ 
 X-2 Alkali & AVCO-RAD & 944 & 123 & \cite{MPDCollection} \\
 LAJ-AF-2 & EOS & 759 & 21 & \cite{MPDCollection} \\ 
 130 kW Lithium MPD  & RIAME MAI & 145 & 130 & \cite{Tikhonov_MPD} \\ 
 X16 & Stuttgart/DFVLR & 120 & 15 & \cite{MPDCollection} \\
 Li LFA & Princeton University & 116 & 100 & \cite{Choueiri1996} \\
 Tokyo10kW & Tokyo University & 32 & 7 & \cite{MPDCollection} \\
 \hline
 Hall Thruster & Organization & Thrust Density & Power & Reference \\ 
  &  & (N/m$^2$) & (kW) & \\ 
  \hline
 SPT-290 & Fakel & 22.7 & 30 & \cite{jankovsky_high_1999} \\ 
 BHT-8000 & Busek & 22.3 & 8 & \cite{pote_performance_1999} \\ 
 SPT-140 & Fakel & 18.6 & 4.6 & \cite{manzella_performance_1997} \\
 NASA-400M & NASA & 16.7 & 47 & \cite{jacobson_nasas_2004} \\ 
  NASA-300M & NASA & 16.0 & 20 & \cite{kamhawi_performance_2013} \\
  SPT-200 & Fakel & 15.8 & 11 & \cite{jankovsky_high_1999} \\ 
  BHT-20k & Busek & 14.1 & 20 & \cite{szabo_commercial_2011} \\ 
  NASA-457Mv2 & NASA & 14.0 & 50 & \cite{soulas_performance_2012} \\ 
 T-220 & NASA & 13.2 & 10 & \cite{mason_1000_2001} \\ 
 PPS-20k & Snecma & 13.1 & 23.5 & \cite{zurbach_performance_2011} \\ 
 PPS-1350 & Snecma & 12.7 & 1.8 & \cite{albarede_characteristics_2005} \\ 
 BHT-600 & Busek & 12.1 & 0.8 & \cite{szabo_one_2020} \\ 
 SPT-100 & Fakel & 11.5 & 1.4 & \cite{sankovic_performance_1994} \\
 CHTpm & PPPL & 10.6 & 0.18 & \cite{polzin_comparisons_2010} \\ 
 BHT-1500 & Busek & 9.1 & 1.8 & \cite{diamant_performance_2015} \\ 
 \hline
 Ion Thruster & Organization & Thrust Density & Power & Reference \\
    &  & (N/m$^2$) & (W) & \\
 \hline
 XIPS 25cm & L3 Technologies & 3.4 & 4.3 & \cite{GoebelBook} \\ 
 NEXT & NASA & 2.3 & 6.9 & \cite{Anderson2000PerformanceCO} \\ 
 T5 Kaufman & Mitsubishi & 2.3 & 0.5 & \cite{GoebelBook} \\ 
 ETS-8 Kaufman & Mitsubishi & 2.1 & 0.6 & \cite{GoebelBook} \\ 
 RIT-RD & Astrium & 1.9 & 0.5 & \cite{GoebelBook} \\ 
 NSTAR & NASA & 1.3 & 2.3 & \cite{NEXT_Performance} \\ 
 XIPS 10cm & L3 Technologies & 1.3 & 0.4 & \cite{GoebelBook} \\ 
 µ10 ECR & JAXA & 0.1 & 0.3 & \cite{GoebelBook} \\ 
 \hline
 Electrospray Thruster & Organization & Thrust Density & Power & Reference \\ 
   &  & (N/m$^2$) & (W) & \\
 \hline
 Indium MEP 2016 & JPL & 0.29 & 3.5 & \cite{Indium_MEP} \\ 
 BET-1mN & Busek & 0.097 & 15 & \cite{Busek_BET1mN} \\ 
 S-iEPS thruster & NASA/MIT & 0.06 & 0.15 & \cite{Electrospray_Scaling} \\ 
 TILE & NASA/MIT & 0.058 & 10 & \cite{Electrospray_Scaling} \\ 
  TILE 3 & Accion & 0.045 & 20 & \cite{AccionTile3} \\ 
 BET-100 & Busek & 0.012 & 5.5 & \cite{Busek_BET100} \\ 
 
 \hline
\end{tabular}
\label{table:thrustdensreferences}
\end{table}

\section{Magnetic Curvature}\label{sec:AppendixCurve}

The radius of curvature for a given field falls out of the tension term of a field line:

\begin{equation}
    (\vec B \cdot \vec \nabla ) \vec B = B \frac{d(B \hat s)}{ds} = B^2 \frac{d(\hat s)}{ds} + B\frac{dB}{ds} \hat s = B^2 \frac{\hat n}{R_B} + \hat s \frac{d(B^2)}{2 ds}\,.
\end{equation}

The radius of curvature can then be written, in the common inverse form, as the amplitude of the normal component $\hat n$:

\begin{equation}
    \frac{1}{R_B} = \frac{\norm{(\vec B \cdot \vec \nabla)\vec B - \hat s \frac{d(B^2)}{2ds}}}{\vec B \cdot \vec B}\,.
\end{equation}

To determine the total radius of curvature for the superposition of two fields, such as the induced magnetic field and the existing external field, we can express the total magnetic field as the sum of these two fields: $\vec B = \vec B_{ind} + \vec B_{ext}$\,.

\begin{equation}
\begin{aligned}
    \frac{1}{R_B} &= \frac{\norm{((\vec B_{ext} + \vec B_{ind}) \cdot \vec \nabla) (\vec B_{ext} + \vec B_{ind}) - \hat s \frac{d(B^2)}{2ds}}}{(\vec B_{ext} + \vec B_{ind}) \cdot (\vec B_{ext} + \vec B_{ind})} \\
    &= \frac{\norm{(\vec B_{ind} \cdot \vec \nabla) \vec B_{ind} + (\vec B_{ind} \cdot \vec \nabla) \vec B_{ext} + (\vec B_{ext} \cdot \vec \nabla) \vec B_{ind} + (\vec B_{ext} \cdot \vec \nabla) \vec B_{ext} - \hat s \frac{d(B^2)}{2ds}}}{(\vec B_{ext} + \vec B_{ind}) \cdot (\vec B_{ext} + \vec B_{ind})}\,. \\
    \end{aligned}
\end{equation}

The final term on the right hand side can be expanded to:

\begin{equation}\label{eq:totalcurve1}
\begin{aligned}
    - \hat s \frac{d(B^2)}{2ds} &= - \hat s \frac{d((\vec B_{ext} + \vec B_{ind}) \cdot (\vec B_{ext} + \vec B_{ind}))}{2ds} \\
    &=- \hat s \frac{d(\vec B_{ind} \cdot \vec B_{ind} + \vec B_{ext} \cdot \vec B_{ext} + 2\vec B_{ext} \cdot \vec B_{ind})}{2ds} \\
    &= - \hat s \frac{d(B_{ind}^2+B_{ext}^2 + 2\vec B_{ext} \cdot \vec B_{ind})}{2ds}\,.
\end{aligned}
\end{equation}

The radius of curvature for the induced and external fields can be substituted into Eq. (\ref{eq:totalcurve1}) to simplify, where it is noted that along the integration path $\vec B_{ind}$ and $\vec B_{ext}$ are parallel and the radii of curvature vector for the induced and external field are in the same direction:

\begin{equation}\label{eq:totalcurve2}
\begin{aligned}
    \frac{1}{R_B} &= \Big\Vert\left((\vec B_{ind} \cdot \vec \nabla) \vec B_{ind} - \hat s \frac{d(B_{ind}^2)}{2ds}\right) + \left((\vec B_{ext} \cdot \vec \nabla) \vec B_{ext} - \hat s \frac{d(B_{ext}^2)}{2ds}\right)\\
    &+ (\vec B_{ind} \cdot \vec \nabla) \vec B_{ext} + (\vec B_{ext} \cdot \vec \nabla) \vec B_{ind} \\
    &- \hat s \frac{d(2\vec B_{ext} \cdot \vec B_{ind})}{2ds}\Big\Vert / \left( {(\vec B_{ext} + \vec B_{ind}) \cdot (\vec B_{ext} + \vec B_{ind})}\right) \\
    &= \frac{\norm{ \frac{\hat n B_{ind}^2}{R_{B,ind}} + \frac{\hat n B_{ext}^2}{R_{B,ext}}+(\vec B_{ind} \cdot \vec \nabla) \vec B_{ext} + (\vec B_{ext} \cdot \vec \nabla) \vec B_{ind}  - \hat s \frac{d(\vec B_{ext} \cdot \vec B_{ind})}{ds}}}{(\vec B_{ext} + \vec B_{ind}) \cdot (\vec B_{ext} + \vec B_{ind})}\,.
\end{aligned}
\end{equation}

This equation can be further simplified by finding the change of $\vec B_{ext}$ along $\vec B_{ind}$, noting that along the integration path $\hat s_{ext} = \hat s = \pm \hat s_{ind}$, where the induced field is in the same direction below the current tube and in the opposite direction above the current tube:

\begin{equation}
    \begin{aligned}
        (\vec B_{ind} \cdot \vec \nabla) \vec B_{ext} &= B_{ind} \frac{d(B_{ext}\hat s_{ext})}{ds_{ind}}\\
        &= B_{ind}B_{ext}\frac{d \hat s_{ext}}{ds_{ind}} + B_{ind}\hat s_{ext}\frac{dB_{ext}}{ds_{ind}}\\
        &=(\hat s_{ind} \cdot \hat s_{ext})\left(B_{ind}B_{ext}\frac{\hat n}{R_{B,ext}} + \hat s\frac{d(B_{ext}B_{ind})}{ds} - B_{ext}\hat s \frac{dB_{ind}}{ds}\right)\,.
    \end{aligned}
\end{equation}

The change of $\vec B_{ind}$ along $\vec B_{ext}$ can be found in the same manner, which provides a sum of: 

\begin{equation}
\begin{aligned}
    (\vec B_{ind} \cdot \vec \nabla) \vec B_{ext} + (\vec B_{ext} \cdot \vec \nabla) \vec B_{ind} &= (\hat s_{ind} \cdot \hat s_{ext}) \left( B_{ind}B_{ext}\frac{\hat n}{R_{B,ext}} + \hat s\frac{d(B_{ext}B_{ind})}{ds}\right.\\
    &- B_{ext}\hat s \frac{dB_{ind}}{ds} + \left.B_{ind}B_{ext}\frac{\hat n}{R_{B,ind}} + B_{ext}\hat s\frac{d(B_{ind})}{ds}\right) \\
    &=(\hat s_{ind} \cdot \hat s_{ext}) \left(B_{ind}B_{ext}\left(\frac{\hat n}{R_{B,ext}}+\frac{\hat n}{R_{B,ind}}\right)\right. \\
    &+\left. \hat s\frac{d(B_{ext}B_{ind})}{ds}\right)
    \end{aligned}
\end{equation}
This can then be substituted into Eq. (\ref{eq:totalcurve2}), noting that the radius of curvature is defined by the amplitude of the normal component:

\begin{equation}
    \frac{1}{R_B} = \frac{\frac{B_{ind}^2}{R_{B,ind}} + \frac{B_{ext}^2}{R_{B,ext}} + (\hat s_{ind} \cdot \hat s_{ext}) B_{ind}B_{ext}\left(\frac{1}{R_{B,ext}}+\frac{1}{R_{B,ind}}\right)}{(\vec B_{ext} + \vec B_{ind}) \cdot (\vec B_{ext} + B_{ind})}\,.
\end{equation}

The bottom ($R_{bot}$) and the top ($R_{top}$) radii of curvature are finally:

\begin{equation}\label{eq:Rtop}
    \frac{1}{R_{top}} = \frac{\frac{B^2_{ext}}{R_{B,ext}} + \frac{B^2_{ind}}{R_{B,ind}} - B_{ind}B_{ext}\left(\frac{1}{R_{B,ind}} + \frac{1}{R_{B,ext}}\right)}{(B_{ext}-B_{ind})^2}\,,
\end{equation}

\begin{equation}\label{eq:Rbot}
    \frac{1}{R_{bot}} = \frac{\frac{B^2_{ext}}{R_{B,ext}} + \frac{B^2_{ind}}{R_{B,ind}} + B_{ind}B_{ext}\left(\frac{1}{R_{B,ind}} + \frac{1}{R_{B,ext}}\right)}{(B_{ext}+B_{ind})^2}\,.
\end{equation}

We can substitute the radii of curvature of Eq. (\ref{eq:Rtop}) and Eq. (\ref{eq:Rbot}) into Eq. (\ref{eq:magpressurederive}) to integrate the magnetic tension term over the radius of the current tube $R$, noting that the orientation of the radius of curvature is negative from $z_0$ to $0$ and positive from $0$ to $z_f$ (Fig. \ref{fig:annularsketch}) and the radius of curvature of the induced field is simply the distance from the center of the current tube $z$:

\begin{equation}\label{eq:longmagpressderivecurve}
\begin{aligned}
    \mathcal T_1 &= \frac{4 B_{IND} B_{ext}}{2 \mu_0} + \int_{0}^{R}{\left(\frac{(B_{ext} + B_{ind})^2}{\mu_0 R_B}  \right) dz} - \int_{0}^{R}{\left(\frac{(B_{ext} - B_{ind})^2}{\mu_0 R_B}  \right) dz} \\
    &= \frac{4 B_{IND} B_{ext}}{2 \mu_0} + \frac{1}{\mu_0}\int_{0}^{R}{\left(\frac{B^2_{ext}}{R_{B,ext}} + \frac{B^2_{ind}}{R_{B,ind}} + B_{ind}B_{ext}\left(\frac{1}{R_{B,ind}} + \frac{1}{R_{B,ext}}\right) \right)dz} \\
    &- \frac{1}{\mu_0}\int_{0}^{R}{\left(\frac{B^2_{ext}}{R_{B,ext}} + \frac{B^2_{ind}}{R_{B,ind}} - B_{ind}B_{ext}\left(\frac{1}{R_{B,ind}} + \frac{1}{R_{B,ext}}\right) \right)dz} \\
    &= \frac{4 B_{IND} B_{ext}}{2 \mu_0} + \frac{2}{\mu_0}\int_{0}^{R}{\left(B_{ind}B_{ext}\left(\frac{1}{z} + \frac{1}{R_{B,ext}}\right) \right)dz} \\
    &= \frac{4 B_{IND} B_{ext}}{2 \mu_0} + \frac{2}{\mu_0}\int_{0}^{R}{\left(\frac{\mu_0 j z}{2} B_{ext}\left(\frac{1}{z} + \frac{1}{R_{B,ext}}\right) \right)dz} \\
    &= \frac{4 B_{IND} B_{ext}}{2 \mu_0} + \frac{2}{\mu_0}\frac{\mu_0 j}{2}B_{ext}\int_{0}^{R}{\left(1 + \frac{z}{R_{B,ext}}\right)dz} \\
    &= \frac{4 B_{IND} B_{ext}}{2 \mu_0} + \frac{2}{\mu_0}\frac{\mu_0 j}{2}B_{ext}\left(R + \frac{R^2}{2 R_{B,ext}}\right) \\
    &= \frac{4 B_{IND} B_{ext}}{2 \mu_0} + \frac{4 B_{IND} B_{ext}}{2 \mu_0}\left(1 + \frac{R}{2 R_{B,ext}}\right)\,. \\
    \end{aligned}
\end{equation}

If we assume the radius of curvature of the externally applied magnetic field is very large compared to the radius of the current tube ($R_{ext}>>R$), which we expect for straight radial magnetic fields and as is shown in the main text, Eq. (\ref{eq:longmagpressderivecurve}) simplifies to the following:
\begin{equation}\label{eq:magpressurethrustdenscurve}
    \mathcal T_1 \approx \frac{4 B_{IND} B_{ext}}{2 \mu_0} + \frac{4 B_{IND} B_{ext}}{2 \mu_0}\,.
\end{equation}

However it is noted that a lower radius of curvature on the externally applied field provides an increase to the thrust density by this additional term $\frac{4 B_{IND} B_{ext}}{2 \mu_0}\frac{R}{2 R_{B,ext}}$.

\section{Minimum Hall Parameter}\label{sec:AppendixHall}

The value of the Hall parameter should correspond to the condition where the electrons display closed drift behavior. That is, the electrons complete a rotation around the thruster channel faster than they move through the acceleration region. This can be written in terms of an inequality:

\begin{equation}
    \frac{v_{\theta}}{2 \pi R_{chan}} > \frac{v_{z}}{L_{acc}}\,,
\end{equation}
where we have defined the azimuthal electron speed $v_{\theta}$, the thruster channel radius $R_{chan}$, the electrons to move axially with speed $v_{z}$ and the acceleration region length to be $L_{acc}$. In our idealized model of the Hall thruster where the electrons are focused in a tube of radius $R$, if we assume the voltage drop is localized in this tube then the length of the acceleration region corresponds to the diameter of the tube; $L_{acc} = 2R$. Noting that the axial electron velocity is due to the drift of electrons by frequency $\nu$ and gyroradius $r_{Le}$:

\begin{equation}
    \frac{v_{\theta}}{2 \pi R_{chan}} > \nu \frac{r_{Le}}{2R}\,,
\end{equation}
By taking the azimuthal electron velocity to be due to $E \times B$ drift, we can transform this inequality into one with the Hall parameter, where we take the electron velocity in gyromotion to be the thermal electron speed:

\begin{equation}
    \begin{aligned}
        \frac{v_{\theta}}{2 \pi R_{chan}} &> \nu \frac{r_{Le}}{2R}\,,\\
        \frac{E}{2 \pi R_{chan} B_{ext}} &> \nu \frac{v_e m_e }{2R e B_{ext}}\,, \\
        \frac{V_D/2R}{2 \pi R_{chan} B_{ext} \nu} &>  \frac{m_e }{2R e B_{ext}}\sqrt{\frac{T_e}{m_e}}\,, \\
        \frac{V_D/2R e B m_e}{2 \pi R_{chan} B_{ext}^2 e m_e \nu} &>  \frac{m_e }{2R e B_{ext}}\sqrt{\frac{T_e}{m_e}}\,, \\
        \frac{e B_ext }{m_e \nu} &>  \frac{2 \pi R_{chan} B_{ext} }{V_D} \sqrt{\frac{T_e}{m_e}}\,, \\
        \Omega_{H} &>  \frac{2 \pi R_{chan} B_{ext} }{V_D} \sqrt{\frac{T_e}{m_e}}\,, \\
        \Omega_{H,min} &=  \frac{2 \pi R_{chan} B_{ext} }{V_D} \sqrt{\frac{T_e}{m_e}}\,. \\
    \end{aligned}
\end{equation}

For a typical Hall thruster with $T_e = 30eV$, $V_D=300 V$, $B_{ext}=300 G$, and $R_{chan}= 5 cm$, the $\Omega_{H,min}=72$.

%%=============================================%%
%% For submissions to Nature Portfolio Journals %%
%% please use the heading ``Extended Data''.   %%
%%=============================================%%

%%=============================================================%%
%% Sample for another appendix section			       %%
%%=============================================================%%

%% \section{Example of another appendix section}\label{secA2}%
%% Appendices may be used for helpful, supporting or essential material that would otherwise 
%% clutter, break up or be distracting to the text. Appendices can consist of sections, figures, 
%% tables and equations etc.

\end{appendices}

%%===========================================================================================%%
%% If you are submitting to one of the Nature Portfolio journals, using the eJP submission   %%
%% system, please include the references within the manuscript file itself. You may do this  %%
%% by copying the reference list from your .bbl file, paste it into the main manuscript .tex %%
%% file, and delete the associated \verb+\bibliography+ commands.                            %%
%%===========================================================================================%%

\bibliography{sn-bibliography}% common bib file
%% if required, the content of .bbl file can be included here once bbl is generated
%%\input sn-article.bbl

%% Default %%
%%\input sn-sample-bib.tex%

\end{document}